\documentclass[journal, 10pt]{IEEEtran}

\usepackage{cite}
\usepackage{amsmath,amssymb,amsfonts,amsthm, empheq}
\usepackage{algorithm}
\usepackage{algorithmic}
\usepackage{graphicx}
\usepackage{textcomp}
\usepackage{xcolor}
\usepackage{mathtools}
\usepackage{graphicx}
\usepackage{booktabs}
\usepackage{makecell}
\usepackage[font={small}]{caption}
\usepackage{subcaption}
\usepackage{cite}
\usepackage{breqn}
\usepackage{mathrsfs}
\usepackage{accents}
\usepackage{acronym}
\usepackage{multirow}
\usepackage{soul}
\usepackage{bm}
\usepackage{url}
\usepackage{wrapfig}
\usepackage{todonotes}
\usepackage{verbatim}
\usepackage{pdfpages}
\usepackage{siunitx}

\usepackage{pgfplots}
\pgfplotsset{compat=1.3}
\usepgfplotslibrary{statistics}
\usepackage{tikz}
\usepackage{pgfplotstable}
\usetikzlibrary{shapes, arrows, positioning, calc, backgrounds, angles, quotes, patterns}

\renewcommand{\thetable}{\arabic{table}}

\DeclareMathOperator*{\argmin}{arg\,min}

\graphicspath{{./figures/}}
\acrodef{prop}[\textit{MIMORPH}]{MIMO Radio Platform for Heterogeneous wireless systems}

\acrodef{abft}[A-BFT]{Association Beamforming Training}
\acrodef{ack}[ACK]{Acknowledge}
\acrodef{adc}[ADC]{Analog-to-Digital Converter}
\acrodef{aoa}[AoA]{Angle of Arrival}
\acrodef{aod}[AoD]{Angle of Departure}
\acrodef{ap}[AP]{Access Point}
\acrodef{amc}[AMC]{Advanced Mezzanine Card}
\acrodef{awv}[AWV]{Antenna Wave Vector}
\acrodef{axi}[AXI]{Advanced eXtensible Interface}
\acrodef{ber}[BER]{Bit Error Rate}
\acrodef{bft}[BFT]{Beamforming Training}
\acrodef{bp}[BP]{Beam Pattern}
\acrodef{brp}[BRP]{Beam Refinement Phase}
\acrodef{cs}[CS]{Compressed Sensing}
\acrodef{cdf}[CDF]{Cumulative Distribution Function}
\acrodef{cef}[CEF]{Channel Estimation Field}
\acrodef{cfo}[CFO]{Carrier Frequency Offset}
\acrodef{cir}[CIR]{Channel Impulse Response}
\acrodef{cfr}[CFR]{Channel Frequency Response}
\acrodef{csi}[CSI]{Channel State Information}
\acrodef{csirs}[CSI-RS]{CSI-Reference Signal}
\acrodef{cs}[CS]{Compressed Sensing}
\acrodef{cv}[CV]{Constant Velocity}
\acrodef{cnn}[CNN]{Convolutional Neural Network}
\acrodef{cots}[COTS]{Commercial-Off-The-Shelf}
\acrodef{dft}[DFT]{Discrete Fourier Transform}
\acrodef{dl}[DL]{Deep Learning}
\acrodef{dma}[DMA]{Direct Memory Access}
\acrodef{dmg}[DMG]{Directional Multi Gigabit}
\acrodef{dti}[DTI]{Data Transfer Interval}
\acrodef{edmg}[EDMG]{Enhanced Directional Multi Gigabit}
\acrodef{ekf}[EKF]{Extended Kalman Filter}
\acrodef{elu}[ELU]{Exponential-Linear Unit}
\acrodef{fmcw}[FMCW]{Frequency-Modulated Continuous-Wave}
\acrodef{fov}[FOV]{Field-of-View}
\acrodef{ft}[FT]{Fourier Transform}
\acrodef{fr2}[FR2]{Frequency Range 2}
\acrodef{gpio}[GPIO]{General Purpose Input/Output}
\acrodef{gsps}[GSPS]{Giga-Samples per Second}
\acrodef{har}[HAR]{Human Activity Recognition}
\acrodef{ht}[HT]{High Throughput}
\acrodef{idft}[IDFT]{Inverse Discrete Fourier Transform}
\acrodef{if}[IF]{Intermediate Frequency}
\acrodef{ifs}[IFS]{Inter-Frame Spacing}
\acrodef{iht}[IHT]{Iterative Hard Thresholding}
\acrodef{ista}[ISTA]{Iterative Shrinkage-Thresholding Algorithm}
\acrodef{isac}[ISAC]{Integrated Sensing And Communication}
\acrodef{imu}[IMU]{Inertial Measurement Unit}
\acrodef{jcs}[JCS]{Joint Communication and Sensing}
\acrodef{jpdaf}[JPDAF]{Joint Probabilistic Data Association Filter}
\acrodef{los}[LoS]{Line-of-Sight}
\acrodef{lbm}[LBM]{Loop-Back Memory}
\acrodef{mae}[MAE]{Mean Absolute Error}
\acrodef{mcs}[MCS]{Modulation and Coding Scheme}
\acrodef{md}[$\mu$D]{micro-Doppler}
\acrodef{mimo}[MIMO]{Multiple Input Multiple Output}
\acrodef{mmwave}[mmWave]{Millimeter-Wave}
\acrodef{msps}[MSPS]{Mega-Samples per Second}
\acrodef{mu}[MU]{Multiple User}
\acrodef{MUSIC}[MUSIC]{MUlti SIgnal Classification}
\acrodef{nac}[NAC]{Normalized Auto Correlation}
\acrodef{nco}[NCO]{Numerical Controlled Oscillator}
\acrodef{nlos}[NLoS]{Non-Line-of-Sight}
\acrodef{nn}[NN]{Neural Network}
\acrodef{nls}[NLS]{Nonlinear Least-Squares}
\acrodef{ofdm}[OFDM]{Orthogonal Frequency Division Multiplexing}
\acrodef{omp}[OMP]{Orthogonal Matching Pursuit}
\acrodef{per}[PER]{Packet Error Rate}
\acrodef{phy}[PHY]{Physical Layer}
\acrodef{pl}[PL]{Programmable Logic}
\acrodef{pov}[POV]{Point-of-View}
\acrodef{ps}[PS]{Processing System}
\acrodef{po}[PO]{Phase Offset}
\acrodef{qam}[QAM]{Quadrature Amplitude Modulation}
\acrodef{ransac}[RANSAC]{Random Sample Consensus}
\acrodef{rf}[RF]{Radio Frequency}
\acrodef{rfsoc}[RFSoC]{Radio Frequency System on a Chip}
\acrodef{rcs}[RCS]{Radar Cross-Section}
\acrodef{rss}[RSS]{Received Signal Strength}
\acrodef{rom}[ROM]{Read Only Memories}
\acrodef{rx}[RX]{receiver}
\acrodef{sc}[SC]{Single Carrier}
\acrodef{sdr}[SDR]{Software Defined Radio}
\acrodef{siso}[SISO]{Single Input Single Output}
\acrodef{sls}[SLS]{Sector Level Sweep}
\acrodef{snr}[SNR]{Signal-to-Noise Ratio}
\acrodef{soc}[SoC]{System on a Chip}
\acrodef{spb}[SPB]{Signal Processing Blocks}
\acrodef{srrc}[SRRC]{Square-Root-Raised-Cosine}
\acrodef{ssb}[SSB]{Synchronization Signal Block}
\acrodef{ssr}[SSR]{Super Sample Rate}
\acrodef{sta}[STA]{Station}
\acrodef{std}[STD]{Standard Deviation}
\acrodef{stf}[STF]{Short Training Field}
\acrodef{stft}[STFT]{Short Time Fourier Transform}
\acrodef{su}[SU]{Single User}
\acrodef{tf}[TF]{Time-Frequency}
\acrodef{to}[TO]{Timing Offset}
\acrodef{toa}[ToA]{Time of Arrival}
\acrodef{tx}[TX]{transmitter}
\acrodef{ue}[UE]{User Equipment}
\acrodef{ula}[ULA]{Uniform Linear Array}
\acrodef{usrp}[USRP]{Universal Software Radio Peripheral}
\acrodef{vht}[VHT]{Very High Throughput}
\acrodef{wlan}[WLAN]{Wireless Local Area Network}

\newcommand{\eq}[1]{Eq.~\eqref{#1}}
\newcommand{\eqs}[2]{Eqs.~(\ref{#1}-\ref{#2})}
\newcommand{\fig}[1]{Fig.~\ref{#1}}
\newcommand{\tab}[1]{Tab.~\ref{#1}}
\newcommand{\secref}[1]{Section~\ref{#1}}

\newcommand\rev[1]{\textcolor{black}{#1}}

\newcommand{\mytexttilde}{{\raise.17ex\hbox{$\scriptstyle\mathtt{\sim}$}}}

\IEEEaftertitletext{\vspace{-2.5\baselineskip}}

\begin{document}

\title{\LARGE Bistatic Doppler Frequency Estimation with Asynchronous \\Moving Devices for Integrated Sensing and Communications}

\author{Gianmaria Ventura$^{\dag}$,~\IEEEmembership{Graduate Student Member,~IEEE}, Zaman Bhalli$^{\dag}$,~\IEEEmembership{Graduate Student Member,~IEEE}, \\Michele Rossi$^{\dag *}$~\IEEEmembership{Senior Member,~IEEE}, Jacopo Pegoraro$^{\dag}$,~\IEEEmembership{Member,~IEEE}
\thanks{$^{\dag}$These authors are with the University of Padova, Department of Information Engineering. $^{*}$This author is with the University of Padova, Department of Mathematics.
Corresponding author email: \texttt{gianmaria.ventura@phd.unipd.it}. 
}
}

\maketitle

\begin{abstract}

In this letter, we present for the first time a method to estimate the bistatic Doppler frequency of a target with \textit{clock asynchronous and mobile} \acf{isac} devices.
Existing approaches have separately tackled the presence of phase offsets due to clock asynchrony \textit{or} the additional Doppler shift due to device movement. However, in real \ac{isac} scenarios, these two sources of phase nuisance are \textit{concurrently} present, making the estimation of the target's Doppler frequency particularly challenging.   
Our method solves the problem using the sole wireless signal at the receiver, exploiting the invariance of phase offsets across multipath components and the bistatic geometry in an original way.
The proposed method is validated via simulation, exploring the impact of different system parameters. Numerical results show that our approach is a viable way of estimating Doppler frequency in bistatic asynchronous \ac{isac} scenarios with mobile devices. 
\end{abstract}

\begin{IEEEkeywords}
Integrated sensing and communication, clock asynchrony, bistatic sensing, mobile devices, Doppler frequency.
\end{IEEEkeywords}

\IEEEpeerreviewmaketitle

\section{Introduction}\label{sec:intro}


In \acf{isac}, estimating the Doppler frequency caused by targets of interest is of key importance.
Besides enhancing the target detection resolution, it enables advanced applications such as target recognition, and human sensing for remote healthcare~\cite{zhang2021enabling, singh2021multi}.

In bistatic \ac{isac}, where the transmitter (TX) and receiver (RX) are spatially separated, the main challenge is the time-varying relative drift between their clocks, which is referred to as \textit{asynchronism}.
Clock asynchronism causes \ac{to}, \ac{cfo}, and a random \ac{po} across different transmissions. Such offsets hinder coherent processing of the channel measurements across time, introducing errors in the estimate of the Doppler frequency~\cite{wu2024sensing}. Moreover, in real \ac{isac} deployments, an additional Doppler frequency shift is caused by the movement of the TX or RX device, which are usually considered to be static in the \ac{isac} literature. While \ac{cfo} is the same for all propagation paths~\cite{pegoraro2023jump}, but changes quickly over time, the device movement evolves slowly but causes a different frequency shift on each propagation path. This makes the combination of Doppler shift due to device movement and \ac{cfo} particularly challenging to compensate for. 
Existing solutions have either focused on \textit{moving} monostatic radar systems, which are not affected by \ac{cfo} and \ac{po}~\cite{zhang2022mobi2sense}, or have tackled asynchronous \ac{isac} systems with \textit{static} devices~\cite{zhao2024multiple, pegoraro2023jump, wu2024sensing}.
Hence, all these approaches are not applicable to realistic scenarios where \ac{isac} devices are both asynchronous \textit{and} mobile. In this case, it is hard to disentangle the Doppler shift due to the movement of devices, the \ac{cfo}, and the target's Doppler frequency.

In this letter, we propose the first approach to estimate the Doppler frequency of a target in bistatic \ac{isac} with asynchronous mobile devices using the sole wireless signal. The key challenge we solve is the superposition of (i)~the target Doppler frequency, (ii)~the TX or RX Doppler frequency, and (iii)~the \ac{cfo} and \ac{po}, where (ii) and (iii) act as a nuisance that hinders the estimation of the target Doppler. The proposed algorithm obtains phase measurements from different multipath reflections in the \acf{cir} across time. Then, it removes phase offsets by 
leveraging the fact that phase offsets are constant for all propagation paths.
In a second step, time-domain phase differences for each propagation path are obtained and used to construct a non-linear system with one equation per multipath component. Different multipath components are affected by the TX or RX motion in different ways, depending on the path geometry. We exploit this as a source of \textit{diversity} to reduce the number of system variables. As a result, the system is solvable if at least $2$ multipath components from static scatterers are available. In this case, the target Doppler frequency can be estimated and refined by aggregating results over a short processing window. 

The main contributions of this letter are:

$1)$ We propose the first solution to the problem of estimating the Doppler frequency of a target in bistatic \ac{isac} systems with mobile asynchronous devices, affected by \ac{cfo} and \ac{po}. 

$2)$ Our solution is based on an original approach that processes \ac{cir} phase measurements across multipath components and subsequent time frames, removing the undesired phase offsets caused by asynchrony and TX or RX movement. 

$3)$ We validate the proposed algorithm via numerical simulation under different system configurations, showing that it provides a viable way to estimate Doppler frequency in realistic bistatic \ac{isac} systems.   

\section{System Model}\label{sec:sysmodel}
This section introduces our system model, including the reference scenario, the \ac{cir}, and the phase measurements.
\vspace{-0.2cm}
\subsection{Reference scenario}
\label{sec:scenario}
We consider a $2$D scenario including two \ac{isac} devices, namely, a transmitter~(TX) and a receiver~(RX). The aim is to estimate the bistatic Doppler frequency caused by a moving target of interest, as shown in \fig{fig:geometric_disp}, by using the channel estimates obtained from the ongoing communication traffic. For simplicity, we assume that the TX is static while the RX moves with velocity $v^{\rm rx}(t)$ with an angle $\eta(t)$ with respect to the segment connecting the TX and the RX (\ac{los} segment), where $t$ is the continuous-time variable. The symmetric case where the TX moves and the RX is static leads to similar derivations, as discussed in \secref{sec:cir-model}, and it is omitted. \rev{Furthermore, the case where both TX and RX move is analogous with $v^{\rm rx}(t)$ being the RX speed relative to the TX.} The received signal includes $M(t)$ delayed, Doppler-shifted, and attenuated copies of the transmitted one, where $M(t)$ corresponds to the number of multipath components caused by scatterers in the environment. As commonly done in \ac{isac}, we only consider first-order reflections since they have significantly higher received power than higher-order ones~\cite{zhang2021overview}. 
Denote by \mbox{$\lambda = c/f_{\rm c}$} the transmission wavelength, with $c$ being the speed of light and $f_{\rm c}$ the carrier frequency. The movement of the $m$-th scatterer causes a bistatic Doppler shift $f_{{\rm D}, m}(t)$. The RX movement causes a Doppler shift equal to $f_{{\rm D},m}^{\mathrm{rx}}(t) = (v^{\rm rx} (t) \cos \xi_m (t) ) / \lambda $,
where $\xi_m(t)$ is the angle between the elongation of the segment connecting the $m$-th scatterer to the RX and the RX velocity vector. 
In our scenario, we consider that the $M(t)$ propagation paths can be partitioned as follows: (i)~a \ac{los} path, which represents the direct propagation from the TX to the RX (assumed to be available), (ii)~a single moving target path, (iii)~\mbox{$S(t)$} \textit{static} scatterers paths for which \mbox{$f_{{\rm D}, m}(t)=0$}. 
We assume that the orientation of the RX with respect to the TX is known and has been compensated for, e.g., by using onboard sensors like a gyroscope or an \ac{imu}. 
\rev{Note that using such sensors, the RX could also measure $v^{\rm rx}(t)$, which would simplify the derivation of the target's Doppler frequency. However, the accuracy of such measurement is critical to the subsequent estimation process as discussed in~\cite{pegoraro2023jump}. Since low-cost sensors available on commercial cellular devices are often inaccurate, our approach considers $v^{\rm rx}(t)$ to be unknown, i.e., it relies on the sole wireless signal. Still, a reliable estimate of $v^{\rm rx}(t)$ can be readily incorporated into our analysis to either reduce the number of required multipath components or to initialize the \ac{nls} problem of \secref{sec:doppler-est}.}

\vspace{-0.2cm}
\subsection{Channel model}
\label{sec:cir-model}

Consider the continuous-time channel between the TX and the RX and denote by $\tau_m(t)$, $\alpha_m(t)$, and $A_m(t)$ the propagation delay, the \ac{aoa}, and the complex coefficient of the \mbox{$m$-th} path, respectively. $A_m(t)$ accounts for the propagation loss and the complex target reflectivity, hence it contains a phase term that depends on the path.  

The \ac{to}, \ac{cfo}, and random \ac{po} are denoted by $\tau_{\rm o}(t)$, $f_{\rm o}(t)$, and $\psi_{{\rm o}}(t)$, respectively. 
Denoting the Dirac delta function by $\delta_{\tau}$, the \ac{cir} at time $t$ and delay $\tau$ is

\small
\begin{equation}\label{eq:cir-continuous}
    h(t,\tau) = e^{j\psi_{{\rm o}}(t)}\sum_{m=1}^{M(t)}{A_m(t) e^{j\vartheta_m(t)}
    \delta_{\tau-\tau_m(t)-\tau_{\rm o}(t)}},
\end{equation}
\normalsize
with \mbox{$\vartheta_m(t) = 2\pi (f_{{\rm D},m}(t)+f_{\rm o}(t)+f_{{\rm D},m}^{\mathrm{rx}}(t))t$}. Note that $f_{{\rm D},m}(t)$ is summed to the \ac{cfo} and the Doppler shift introduced by the receiver, which act as a nuisance.

In \ac{sc} systems, e.g., IEEE~802.11ay, the RX estimates the \ac{cir} directly using cross-correlation of the received signal with a known pilot sequence. In \ac{ofdm} systems, it can obtain the \ac{cir} via \ac{idft} from an estimate of the \ac{cfr}.  

The \ac{cir} estimation is repeated across multiple frames, indexed by $k$, with period $T$. Using a common assumption in \ac{isac} and radar processing, we consider a short processing window of $K$ frames~\cite{zhang2021overview}, where the parameters, $M(t)$, $\tau_m(t)$, $f_{{\rm D}, m}(t)$, $\alpha_m(t)$, $A_m(t), \eta(t),  \xi_m(t), v^{\rm rx}(t)$, and consequently the RX motion-induced Doppler shift $f_{{\rm D},m}^{\mathrm{rx}}(t)$ can be considered constant. 
Conversely, all the nuisance parameters, $\tau_{\rm o}(t)$, $f_{\rm o}(t)$, $\psi_{{\rm o}}(t)$, are time-varying within the window. \rev{Note that, although for communication the \ac{cfo} is usually considered slowly time-varying, in \ac{isac} we need to consider the \textit{residual} \ac{cfo} after estimation and partial compensation by the receiver. The residual \ac{cfo} is fast time-varying, being the random residual of an imperfect estimation and compensation process~\cite{wu2024sensing}.}

The estimated discrete-time \ac{cir} at time $kT$ is

\small
\begin{equation}
    \label{eq:cir-discrete}
    h[k,l] = e^{j\psi_{{\rm o}}(kT)}\sum_{m=1}^{M}{A_{m}e^{j\vartheta_m[k]}\chi[l-\tau_m-\tau_{\rm o}(kT)]},
\end{equation}
\normalsize
where \mbox{$\vartheta_m[k] = 2\pi (f_{{\rm D},m}+f_{\rm o}(kT)+f_{{\rm D},m}^{\mathrm{rx}})kT$}, $l$ is the discrete delay index, and $M$ is the number of resolvable paths within the window, with $M \leq M(t)$. In \eq{eq:cir-discrete}, $\chi[l]$ replaces the Dirac delta function to account for non-ideal autocorrelation of the pilot sequence (in single carrier systems) or the impact of the finite-length \ac{cfr} estimate (in \ac{ofdm} systems). 
In the following, we focus on estimating the target Doppler frequency under the nuisance due to the \ac{cfo}, \ac{po}, and the RX motion. Hence, we assume that the RX can detect and separate the $M$ multipath components and extract their phase across time. To this end, the \ac{to} can be compensated for by obtaining \textit{relative} delay measurements with respect to the \ac{los}, as done in~\cite{pegoraro2023jump}. 
Our model is agnostic to the number of available antennas at the RX. However, some way of estimating \acp{aoa} of the multipath components is needed, either via digital or analog beamforming.

\begin{figure}
    \centering
    \tikzset{antenna/.style={insert path={-- coordinate (ant#1) ++(0,0.25) -- +(135:0.25) + (0,0) -- +(45:0.25)}}}
\tikzset{station/.style={naming,draw,shape=dart,shape border rotate=90, minimum width=10mm, minimum height=10mm,outer sep=0pt,inner sep=3pt}}
\tikzset{mobile/.style={naming,draw,shape=rectangle,minimum width=12mm,minimum height=6mm, outer sep=0pt,inner sep=3pt}}
\tikzset{naming/.style={align=center,font=\small}}

\newcommand{\BS}[1]{%
\begin{tikzpicture}
\node[station] (base) {#1};

\draw[line join=bevel] (base.100) -- (base.80) -- (base.110) -- (base.70) -- (base.north west) -- (base.north east);
\draw[line join=bevel] (base.100) -- (base.70) (base.110) -- (base.north east);

\draw[line cap=rect] ([yshift=0pt]base.north) [antenna=1];
\end{tikzpicture}
}

\newcommand{\UE}[1]{%
\begin{tikzpicture}[every node/.append style={rectangle,minimum width=0pt}]
\node[mobile] (box) {#1};

\draw ([xshift=.25cm] box.south west) circle (4pt)
      ([xshift=-.25cm]box.south east) circle (4pt);

\fill ([xshift=.25cm] box.south west) circle (1pt)
      ([xshift=-.25cm]box.south east) circle (1pt);

\draw (box.north) [antenna=1];
\end{tikzpicture}
}

\begin{tikzpicture}
    \tikzstyle{disp} = [draw, fill=white, circle, minimum size=0.1em]
    \tikzstyle{target} = [fill, circle] 
    \definecolor{orange}{RGB}{255,158,20}
    \definecolor{green}{RGB}{78,211,78}
    \definecolor{cyan}{RGB}{0,204,204}

    \definecolor{purple1194288}{RGB}{119,42,88}
    \definecolor{brown1814888}{RGB}{181,48,88}
    \definecolor{indianred21811088}{RGB}{218,110,88}
    \definecolor{darkslategray583162}{RGB}{58,31,62}
    \definecolor{burlywood231175145}{RGB}{231,175,145}
    
    \begin{scope}
        \node (bs) {\BS{TX}};
        \node [] at(bs.north) (tx) {};

        \node [target, below=4em of tx, xshift=4em](target) {};
        \node[below] at (target.south) {Target};
        \node [right=16.5em of tx, anchor=north, yshift=-2em](ue) {\UE{RX}};
        \node [] at(ue.north) (rx) {};
        
        \draw [-, ultra thick] ($(target)+(6.5em,-.5em)$) -- node[yshift=-.5em](middle1){} ($(target)+(8.5em,-.5em)$);
        \node at ($(middle1)-(0em,.2em)$) () {Static path};
        
        \node [darkslategray583162](v) at ($(rx)+(-4.5em,3em)$) {$v^{\rm rx}(t)$};
        
        \draw [-] (bs.north) -- node[yshift=.5em] () {LoS} (ue.north);
        \node at ($(rx)!-5em!(tx)$)(tempL){};
        \draw [-, dashed] (ue.north) -- (tempL);

        \draw [-] (bs.north) -- (target);
        \draw [-] (target) -- (ue.north);
        \node at ($(rx)!-5em!(target)$)(tempT){};
        \draw [-, dashed] (ue.north) -- (tempT);

        \draw [-] (bs.north) -- ($(middle1)+(0,.5em)$);
        \draw [-] ($(middle1)+(0,.5em)$) -- node[](tempmid1){} (ue.north);
        \node at ($(rx)!-5em!(tempmid1)$)(temp1){};
        \draw [-, dashed] (ue.north) -- (temp1);
        

        \draw [draw=darkslategray583162,->,thick] (ue.north) -- (v);
        
        \pic["$\alpha_{\rm t}$", text=indianred21811088, draw=indianred21811088, -, thick, angle radius=4em, angle eccentricity=.9, pic text options={shift={(-13pt,-2pt)}}] {angle = tx--rx--target};
        \pic[draw=indianred21811088, -, thick, angle radius=3.85em, angle eccentricity=.9] {angle = tx--rx--target};
        \pic["$\xi_{\rm t}$", text=indianred21811088, draw=indianred21811088, -, thick, angle radius=2em, angle eccentricity=.9, pic text options={shift={(30pt, -4pt)}}] {angle = tempT--rx--v};

        \pic["$\alpha_s$", text=brown1814888, draw=brown1814888, -, thick, angle radius=3em, angle eccentricity=.9, pic text options={shift={(-12pt,-10pt)}}] {angle = tx--rx--tempmid1};
        \pic[ draw=brown1814888, -, thick, angle radius=2.85em, angle eccentricity=.9] {angle = tx--rx--tempmid1};
        \pic["$\xi_s$", text=brown1814888, draw=brown1814888, -, thick, angle radius=2.5em, angle eccentricity=.9, pic text options={shift={(1pt,10pt)}}] {angle = temp1--rx--v};


        \pic["$\eta$", xshift=-2em, draw, text=purple1194288, draw=purple1194288, -, thick, angle radius=1.5em, angle eccentricity=.9, pic text options={shift={(23pt,-11.5pt)}}] {angle= tempL--rx--v};
    \end{scope}

\end{tikzpicture}
    \caption{Reference scenario and multipath geometry.}
    \label{fig:geometric_disp}
\end{figure}

\vspace{-0.2cm}
\subsection{Phase measurements}\label{sec:phase-meas}

In this section, we model the phase of each multipath component in the \ac{cir}. 
We denote the noise on \ac{cir} phases at time $k$ due to signal transmission and channel estimation, as $w[k]$.
To simplify the equations, we group the phase nuisance terms, which are the same on all propagation paths, into $\Psi_{{\rm o}}(kT) = \psi_{{\rm o}}(kT) + 2\pi f_{\rm o}(kT)kT$. Moreover, we use subscripts $\cdot_{\rm LoS}$ and $\cdot_{\rm t}$ to refer to quantities related to the \ac{los} and to the target-induced paths, respectively, as shown in~\fig{fig:geometric_disp}. For the paths caused by static scatterers, we use index $s=1, \dots, S$, where $S$ is the number of resolvable static scatterers detected in the processing window, with \mbox{$S = M-2$}.
The phase of the \ac{los} component is

\small
\begin{equation}\label{eq:los-phase}
    \phi_{\rm LoS}[k]= \Psi_{{\rm o}}(kT)+\angle{A_{{\rm LoS}}} + 2\pi  kT\left(\frac{v^{\rm rx}}{\lambda} \cos \eta \right) + w_{\rm LoS}[k],
\end{equation}
\normalsize
where $\angle{\cdot}$ is the phase operator.
The phase of the target-induced path is affected by both the Doppler shift caused by the target, $f_{\rm D, t}$, and by the receiver motion as

\small
\begin{equation}\label{eq:target-phase}
    \phi_{\rm t}[k]=\Psi_{{\rm o}}(kT)+ \angle{A_{{\rm t}}} + 2\pi  kT\left(f_{\rm D, t} + \frac{v^{\rm rx}}{\lambda} \cos \xi_{\rm t} \right) + w_{\rm t}[k],
\end{equation}
\normalsize
where $\xi_{\rm t}$ is the angle between the elongation of the segment connecting the target to the RX and the RX velocity vector. The phase of the $s$-th static multipath component is

\small
\begin{equation}\label{eq:static-phase}
    \phi_{s}[k]=\Psi_{{\rm o}}(kT)+ \angle{A_{s}} + 2\pi  kT \left(\frac{v^{\rm rx}}{\lambda} \cos \xi_s \right) + w_{s}[k].
\end{equation}
\normalsize
 From the \ac{cir}, the RX measures $\texttt{mod}_{2\pi}(\phi_i[k])$ where $i\in\{{\rm t}, {\rm LoS}, s\}$, $k=0, \dots,K-1$, and $\texttt{mod}_{2\pi}(\cdot)$ is the modulo $2\pi$ division, whose result is in $[0, 2\pi]$.

\section{Methodology}\label{sec:method}
In this section, we present our approach for estimating the bistatic Doppler frequency of the target from the phase measurements in \eqs{eq:los-phase}{eq:static-phase}. It can be summarized as follows.

\textit{A. \ac{cfo} and \ac{po} cancellation.} By subtracting the phase of the \ac{los} path from the other phase measurements we \textit{cancel out} the nuisance component $\Psi_{{\rm o}}$, without estimating it directly. This is further detailed in~\secref{sec:cfo-canc}.

\textit{B. Time-domain phase differencing.} By computing time-domain phase differences \textit{for each path}, we cancel out the path-specific phase terms $\angle{A_{i}}$, with $i\in\{{\rm t}, {\rm LoS}, s\}$, as detailed in~\secref{sec:phase-diff}. This yields phase differences whose value depends on the Doppler shifts of the RX and the target.

\textit{C. \ac{aoa}-based simplification.} By leveraging \ac{aoa} estimation at the RX and the multipath geometry, we make a key simplification in the phase measurements model (see~\secref{sec:aoa-simp}). This allows reducing the number of unknowns, making the estimation of the target Doppler frequency feasible if at least $2$ static multipath components are detected.  

\textit{D. Doppler frequency estimation.} We formulate the estimation of the target's Doppler frequency as a \ac{nls} problem across the multipath components (see \secref{sec:doppler-est}). A closed-form solution using $S=2$ static paths is provided next to initialize the \ac{nls} problem.

\vspace{-0.2cm}
\subsection{\ac{cfo} and \ac{po} cancellation}\label{sec:cfo-canc}

We subtract $\phi_{\rm LoS}[k]$ from $\phi_{\rm t}[k]$ and $\phi_s[k]$, obtaining

\small
\begin{align}
        & \tilde{\phi}_{\rm t}[k]= \Omega_{{\rm t}} + 2\pi kT\left(f_{\rm D, t} + \frac{v^{\rm rx}}{\lambda} \left(\cos \xi_{\rm t} - \cos \eta \right) \right)  + w'_{\rm t}[k], \label{eq:new-target-phase} \\ 
        & \tilde{\phi}_{s}[k]= \Omega_{s} + 2\pi kT\left(\frac{v^{\rm rx}}{\lambda} \left(\cos \xi_s - \cos \eta\right)\right) + w'_{s}[k], \label{eq:new-static-phase}
\end{align}
\normalsize

where \mbox{$\Omega_{s} = \angle{A_{s}} - \angle{A_{{\rm LoS}}}$},  \mbox{$\Omega_{{\rm t}} = \angle{A_{{\rm t}}} - \angle{A_{{\rm LoS}}}$}, and $w'_i[k]=w_i[k] - w_{\rm LoS}[k], i\in\{{\rm t}, s\}$. 
Computing phase differences \textit{cancels out} \ac{cfo} and \ac{po} without estimating them, since they are common to all propagation paths.
Despite the absence of \ac{cfo} and \ac{po} in \eq{eq:new-target-phase}, the estimation of the target's Doppler frequency remains non-trivial due to the presence of the undesired frequency term $v^{\rm rx} \left(\cos \xi_{\rm t} - \cos \eta \right)/\lambda$  caused by the receiver movement. Indeed, a direct application of standard Fourier-based processing to estimate the Doppler frequency would fail to separate $f_{\rm D, t}$ from the Doppler due to the RX movement. 

\vspace{-0.2cm}
\subsection{Time-domain phase differencing}\label{sec:phase-diff}

After \ac{cfo} and \ac{po} cancellation, the RX computes first-order, time-domain phase differences as \mbox{$\Delta_i[k] = \texttt{mod}_{2\pi}(\tilde{\phi}_i[k]) - \texttt{mod}_{2\pi}(\tilde{\phi}_i[k-1])$} with $i\in\{{\rm t}, s\}$ and $k = 1, \dots ,K-1$. 
The phase differences are expressed as

\small
\begin{align}
    & \Delta_{\rm t}[k] = 2\pi T\left(f_{\rm D, t}+\frac{v^{\rm rx}}{\lambda}\left( \cos \xi_{\rm t} - \cos \eta\right) \right)+ w''_{\rm t}[k], \label{eq:phi-trg} \\
    & \Delta_s[k] = 2\pi T\left(\frac{v^{\rm rx}}{\lambda} \left(\cos \xi_s - \cos \eta \right)\right) + w''_s[k], \label{eq:phi-m}  
\end{align}
\normalsize
where $w''_i[k]=w'_i[k] - w'_{i}[k-1],i\in\{{\rm t}, s\}$.

In \eq{eq:phi-trg} and \eq{eq:phi-m}, we assume that the channel estimation period $T$ is sufficiently small, so that the phase change between two subsequent frames is smaller than $\pi$. This allows writing phase differences without the ambiguity due to the $\texttt{mod}_{2\pi}$ operator. From \eq{eq:phi-trg}, it can be seen that the noise-free phase differences $\Delta_{\rm t}$ and $\Delta_s$ are upper bounded by $ 2\pi T (3f_{\rm max})$, with $i\in\{{\rm t}, s\}$ and $f_{\rm max}$ being the maximum Doppler shift caused by the RX (or the target). $f_{\rm max}$ is a system design parameter that can be set depending on the specific scenario and application.
To fulfill the assumption, it is sufficient to impose \mbox{$|\Delta_i|<\pi$} which yields \mbox{$T<1/(6f_{\rm max})$}. The choice of $T$ is further discussed in \secref{sec:results}.

\vspace{-0.2cm}
\subsection{\ac{aoa}-based simplification}\label{sec:aoa-simp}

By inspecting \fig{fig:geometric_disp}, a key simplification can be made in \eq{eq:phi-trg} and \eq{eq:phi-m} noticing that 
\begin{equation}\label{eq:angle-simple}
    \cos \xi_i = \cos \left(\eta-\alpha_i \right)= \cos \left(\alpha_i-\eta\right), 
\end{equation}
with $i\in \{{\rm t}, s\}$.
This substitution removes the dependency on the unknown and path-dependent angle $\xi_i$. The new dependency on $\alpha_i-\eta$ is easier to handle since the RX estimates $\alpha_i$ and the unknown term $\eta$ is independent of the propagation paths. 
Note that, since $\alpha_s$ and $\alpha_{\rm t}$ are estimated by the RX, this operation reduces the unknowns from $S+4$, i.e., $f_{\rm D, t}, v^{\rm rx}, \eta, \xi_{\rm t}, \xi_1, \dots, \xi_S$, to just $3$, i.e., $f_{\rm D, t}, v^{\rm rx}, \eta$. This makes \eq{eq:phi-trg} and \eq{eq:phi-m}, for $s=1,\dots, S$, a set of $S+1$ equations with $3$ unknowns, which can be solved if the number of static multipath components satisfies $S \geq 2$. In the next section, we provide our solution based on \ac{nls}.

\vspace{-0.2cm}
\subsection{Doppler frequency estimation}
\label{sec:doppler-est}

We reformulate \eq{eq:phi-trg} and \eq{eq:phi-m} in vector notation by introducing: the phase differences vector at time $kT$, $\mathbf{\Delta}[k] = [\Delta_{\rm t}[k], \Delta_{1}[k], \dots, \Delta_{S}[k]]^{\top}$, the unknown parameter vector $\bm{\theta}=[f_{\rm D, t}, \eta, v^{\rm rx}]^{\top}$, the non-linear vector function $\mathbf{g}(\bm{\theta})$, which expresses the non-linear relations in \eq{eq:phi-trg} and \eq{eq:phi-m}. The following model holds $\mathbf{\Delta}[k] = \mathbf{g}(\bm{\theta}) + \mathbf{w}[k]$, where $\mathbf{w}[k]$ 
is the noise vector with components $w''_i[k], i\in\{{\rm t}, 1, \dots, S\}$. 
To reduce the impact of noise, we average the measured phase differences over time obtaining $\Bar{\mathbf{\Delta}} = \sum_{k=1}^{K-1} \mathbf{\Delta}[k] / (K-1)$.


\subsubsection{Nonlinear least-squares solution}\label{sec:nls-sol}

An \ac{nls} problem is solved to retrieve an estimate of the unknown parameters $\hat{\bm{\theta}} = \argmin_{\bm{\theta}} ||\Bar{\mathbf{\Delta}} - \mathbf{g}(\bm{\theta})||_2^2$,
from which we get the Doppler frequency estimate as the first component of the solution, which we denote by $\hat{f}_{\rm D, t}$. This problem can be solved using, e.g., the Levenberg-Marquardt algorithm with a suitable initialization~\cite{nocedal2006numerical}. 
\rev{The computational complexity of the algorithm is $O(d_{\boldsymbol{\theta}}^2(S+1)N_{\rm it})$, where $d_{\boldsymbol{\theta}} = 3$ is the dimension of $\bm{\theta}$ and $N_{\rm it}$ represents the number of iterations needed to converge.}

\subsubsection{Closed form solution} \label{closed-form}

A closed-form solution using $3$ multipath components is used to initialize the \ac{nls}.
Consider \eq{eq:phi-m} with $2$ phase measurements from static paths, i.e., $s=1,2$. Using also \eq{eq:phi-trg}, a system with $3$ equations in $3$ unknowns is attained, which can be solved for $\bm{\theta}$. Denoting by $\bar{\Delta}_{\rm t}$, $\bar{\Delta}_{\rm 1}$, and $\bar{\Delta}_{\rm 2}$ the time-averaged phase differences for the sensing path, and static paths $1$ and $2$, we compute 

\vspace{-0.2cm}

\small
\begin{equation}
    \label{eq:eta}
    \tilde{\eta} = \arctan\left(\frac{ \bar{\Delta}_2 (\cos \alpha_1 - 1) - \bar{\Delta}_1 (\cos \alpha_2 - 1)}{\bar{\Delta}_1 \sin{\alpha_2} - \bar{\Delta}_2 \sin{\alpha_1}}\right).
\end{equation}
\normalsize
Then, $\hat{\eta}$ is obtained as \mbox{$\hat{\eta}=\texttt{mod}_{2\pi}(\tilde{\eta})$} if the argument of the $\arctan$ is positive and \mbox{$\hat{\eta}=\texttt{mod}_{2\pi}(\tilde{\eta} + \pi)$} otherwise.
Eventually the expression of $\hat{f}_{\rm D, t}$ is

\small
\begin{equation}
    \label{eq:fD}
    \hat{f}_{\rm D, t} = \frac{1}{2 \pi T} \left( \bar{\Delta}_{\rm t} - \bar{\Delta}_1 \frac{\cos(\alpha_{\rm t} - \hat{\eta}) - \cos \hat{\eta}}{\cos(\alpha_1-\hat{\eta})-\cos \hat{\eta}}\right).
\end{equation}
\normalsize
The solution requires the following conditions to be met: (i)~$\alpha_i \neq 0, i \in \{1,2,\mathrm{t}\}$, (ii)~$\alpha_{i} \neq \alpha_{\ell}, i \neq \ell \in \{1,2,\mathrm{t}\}$, and (iii)~$\alpha_{i} \neq 2\hat{\eta}, i \in \{1,2\}$. 
Note that violating conditions (i)~and (ii)~ correspond to degenerate scenarios in which the \ac{los} path is not available, or two static multipath components have the same \ac{aoa}.  Condition (iii)~instead is violated if the \ac{aoa} of one of the multipath components is equal to $2\hat{\eta}$.

\begin{figure*} 
    \centering
    \subcaptionbox{$\acs{snr}=\SI{5}{\decibel}$, $\sigma_\alpha=5$°, and $KT=\SI{16}{\milli\second}$.\protect\label{fig:boxplot_path}}[0.33\textwidth]{\begin{tikzpicture}
\definecolor{black28}{RGB}{28,28,28}
\definecolor{brown1814888}{RGB}{181,48,88}
\definecolor{burlywood231175145}{RGB}{231,175,145}
\definecolor{darkgray176}{RGB}{176,176,176}
\definecolor{darkslategray583162}{RGB}{58,31,62}
\definecolor{indianred21811088}{RGB}{218,110,88}
\definecolor{purple1194288}{RGB}{119,42,88}
\matrix[draw, fill=white, ampersand replacement=\&, 
inner sep=2pt, 
    ] at (1.85,2.23) {  
        \node [shape=rectangle, draw=black ,fill=purple1194288, label=right:\scriptsize{$f_{\rm c}=\SI{60}{\giga\hertz}$}] {};\&
        \node [shape=rectangle, draw=black ,fill=brown1814888, label=right:\scriptsize{$f_{\rm c}=\SI{28}{\giga\hertz}$}] {}; \&
        \node [shape=rectangle, draw=black ,fill=indianred21811088, label=right:\scriptsize{$f_{\rm c}=\SI{5}{\giga\hertz}$}] {};\\
    };
  \begin{axis}
    [
    boxplot/draw direction=y,
    ylabel={Error $\varepsilon_{f_{\rm D, t}}$},
    xlabel={No. static paths $S$},
    ylabel shift=-5 pt,
    ymin=-0.011, ymax=0.12,
    ytick={0,0.05,0.1,0.15,0.2,0.25,0.3,0.35,0.4,0.45,0.5},
    yticklabels={$0$,$0.05$,$0.1$,,$0.2$,,$0.3$,,$0.4$,,$0.5$},
    yticklabel style = {/pgf/number format/fixed},
    cycle list={{purple1194288},{brown1814888},{indianred21811088}},
    xtick={1,2,3,4},
    ymajorgrids,
    xticklabels={$2$, $4$, $6$, $8$},
    xlabel style={font=\footnotesize}, ylabel style={font=\footnotesize}, ticklabel style={font=\footnotesize},
    /pgfplots/boxplot/box extend=0.25,
    height=3.5cm,
    width=5.5cm,
    xmin=0.25,
    xmax=4.75
    ]
        \addplot+[
	fill, draw=black,
	boxplot prepared={
		median=0.0088855180658886,
		upper quartile=0.0209539734227628,
		lower quartile=0.00391902827293351,
		upper whisker=0.0464492244710742,
		lower whisker=3.8073139251606e-07,
		draw position=0.7
	},
	] coordinates {};

	\addplot+[
	fill, draw=black,
	boxplot prepared={
		median=0.00718534133437517,
		upper quartile=0.013478554601113,
		lower quartile=0.00323080572839677,
		upper whisker=0.0288460405436409,
		lower whisker=3.72488731602181e-06,
		draw position=1
	},
	] coordinates {};
	
	\addplot+[
	fill, draw=black,
	boxplot prepared={
		median=0.014684166167723,
		upper quartile=0.026866172436961,
		lower quartile=0.00670776891292069,
		upper whisker=0.0570909217257234,
		lower whisker=5.4908094742307e-07,
		draw position=1.3
	},
	] coordinates {};
	
	\addplot+[
	fill, draw=black,
	boxplot prepared={
		median=0.00818442851491701,
		upper quartile=0.0194650498295056,
		lower quartile=0.00360977513993397,
		upper whisker=0.0432063643959109,
		lower whisker=1.81064728527798e-07,
		draw position=1.7
	},
	] coordinates {};

	\addplot+[
	fill, draw=black,
	boxplot prepared={
		median=0.00604477304461719,
		upper quartile=0.0107396339509796,
		lower quartile=0.00284749442855798,
		upper whisker=0.0225597495642557,
		lower whisker=3.96121045367817e-06,
		draw position=2
	},
	] coordinates {};
	
	\addplot+[
	fill, draw=black,
	boxplot prepared={
		median=0.0122240758121243,
		upper quartile=0.0218262751966194,
		lower quartile=0.00564728901485796,
		upper whisker=0.046081459658168,
		lower whisker=2.69779372195922e-06,
		draw position=2.3
	},
	] coordinates {};

	\addplot+[
	fill, draw=black,
	boxplot prepared={
		median=0.00822684762611817,
		upper quartile=0.0193745493950201,
		lower quartile=0.00348781531651318,
		upper whisker=0.043117328559927,
		lower whisker=2.62418187796598e-06,
		draw position=2.7
	},
	] coordinates {};

	\addplot+[
	fill, draw=black,
	boxplot prepared={
		median=0.00572474218285914,
		upper quartile=0.0101082957792132,
		lower quartile=0.00260308700253941,
		upper whisker=0.0213631462133561,
		lower whisker=4.22497369457338e-06,
		draw position=3
	},
	] coordinates {};

	\addplot+[
	fill, draw=black,
	boxplot prepared={
		median=0.0114722664811631,
		upper quartile=0.0203068666023567,
		lower quartile=0.0053815628044971,
		upper whisker=0.0426817464653587,
		lower whisker=6.69389181668977e-07,
		draw position=3.3
	},
	] coordinates {};

	\addplot+[
	fill, draw=black,
	boxplot prepared={
		median=0.00823514915778123,
		upper quartile=0.0199402980292601,
		lower quartile=0.0035599013372046,
		upper whisker=0.0444737941333891,
		lower whisker=7.60817755941828e-08,
		draw position=3.7
	},
	] coordinates {};

	\addplot+[
	fill, draw=black,
	boxplot prepared={
		median=0.00555226176555184,
		upper quartile=0.00982364296308686,
		lower quartile=0.00252441976322741,
		upper whisker=0.020772407587506,
		lower whisker=1.61479239101089e-06,
		draw position=4
	},
	] coordinates {};

	\addplot+[
	fill, draw=black,
	boxplot prepared={
		median=0.0112653129338285,
		upper quartile=0.0197290191729359,
		lower quartile=0.0051512630983051,
		upper whisker=0.0415464821979729,
		lower whisker=5.62320249094955e-07,
		draw position=4.3
	},
	] coordinates {};
    
    \end{axis}

    \begin{axis}[
        xtick = {1,2,3},
        ymin = 0.78,
        ymax = 3.4,
        xmin=0.25,
        xmax=4.75,
        hide x axis,
        hide y axis,
        height=3.5cm,
        width=5.5cm
    ]
        \addplot [draw=black, dashed, mark options={solid, fill=black}, mark=diamond*] 
        coordinates {
            (1, 2.23173)
            (2, 2.37744)
            (3, 2.56007)
            (4, 2.65577)
        };
    \end{axis}

    \begin{axis}[
        ytick = {1,2,3},
        xmin=0.25,
        xmax=4.75,
        ymin = 0.78,
        ymax = 3.4,
        hide x axis,
        axis y line*=right,
        ylabel={\footnotesize{NLS  time [ms]}},
        ylabel shift = -3pt,
        height=3.5cm,
        width=5.5cm,
        ylabel style={font=\footnotesize},
        yticklabel style = {/pgf/number format/fixed, ticklabel style={font=\footnotesize}},
    ]
  \end{axis}
\end{tikzpicture}}
    \subcaptionbox{$\acs{snr}=\SI{5}{\decibel}$, $\sigma_\alpha=5$°, and $S=2$.\protect\label{fig:boxplot_K}}[0.33\textwidth]{\begin{tikzpicture}
\definecolor{black28}{RGB}{28,28,28}
\definecolor{brown1814888}{RGB}{181,48,88}
\definecolor{burlywood231175145}{RGB}{231,175,145}
\definecolor{darkgray176}{RGB}{176,176,176}
\definecolor{darkslategray583162}{RGB}{58,31,62}
\definecolor{indianred21811088}{RGB}{218,110,88}
\definecolor{purple1194288}{RGB}{119,42,88}
\matrix[draw, fill=white, ampersand replacement=\&, 
inner sep=2pt, 
    ] at (1.9,2.23) {  
        \node [shape=rectangle, draw=black ,fill=purple1194288, label=right:\scriptsize{$f_{\rm c}=\SI{60}{\giga\hertz}$}] {};\&
        \node [shape=rectangle, draw=black ,fill=brown1814888, label=right:\scriptsize{$f_{\rm c}=\SI{28}{\giga\hertz}$}] {}; \&
        \node [shape=rectangle, draw=black ,fill=indianred21811088, label=right:\scriptsize{$f_{\rm c}=\SI{5}{\giga\hertz}$}] {};\\
    };
  \begin{axis}
    [
    boxplot/draw direction=y,
    ylabel={Error $\varepsilon_{f_{\rm D, t}}$},
    xlabel={Aggregation window $KT$ [ms]},
    ylabel shift=-5 pt,
    ymin=-0.011, ymax=0.12,
    ytick={0,0.05,0.1,0.15,0.2,0.25,0.3,0.35,0.4,0.45,0.5},
    yticklabels={$0$,$0.05$,$0.1$,,$0.2$,,$0.3$,,$0.4$,,$0.5$},
    yticklabel style = {/pgf/number format/fixed},
    cycle list={{purple1194288},{brown1814888},{indianred21811088}},
    ymajorgrids,
    xtick={1,2,3,4,5,6},
    xticklabels={$2$, $4$, $8$, $16$, $32$, $48$},
    xlabel style={font=\footnotesize}, ylabel style={font=\footnotesize}, ticklabel style={font=\footnotesize},
    /pgfplots/boxplot/box extend=0.25,
    height=3.5cm,
    width=5.5cm,
    ]
    \addplot+[
	fill, draw=black,
	boxplot prepared={
		median=0.0424564500883838,
		upper quartile=0.0958083111110028,
		lower quartile=0.0187214885628071,
		upper whisker=0.21142782165386,
		lower whisker=1.03941433326421e-05,
		draw position=0.7
	},
	] coordinates {};

	\addplot+[
	fill, draw=black,
	boxplot prepared={
		median=0.0607840463303472,
		upper quartile=0.123626211977778,
		lower quartile=0.0268861785691034,
		upper whisker=0.268649790887595,
		lower whisker=4.13435546096529e-06,
		draw position=1
	},
	] coordinates {};

	\addplot+[
	fill, draw=black,
	boxplot prepared={
		median=0.0710226066697776,
		upper quartile=0.136481194599385,
		lower quartile=0.0319147833266328,
		upper whisker=0.293081355462763,
		lower whisker=1.61062838332815e-05,
		draw position=1.3
	},
	] coordinates {};

	\addplot+[
	fill, draw=black,
	boxplot prepared={
		median=0.0237294372422038,
		upper quartile=0.0548705520487249,
		lower quartile=0.0102532572185949,
		upper whisker=0.121783796218399,
		lower whisker=7.54677180117112e-06,
		draw position=1.7
	},
	] coordinates {};

	\addplot+[
	fill, draw=black,
	boxplot prepared={
		median=0.0304395366254032,
		upper quartile=0.0590800510343915,
		lower quartile=0.0139082925621113,
		upper whisker=0.126801665926237,
		lower whisker=9.25087152787472e-06,
		draw position=2
	},
	] coordinates {};

	\addplot+[
	fill, draw=black,
	boxplot prepared={
		median=0.0347098844270958,
		upper quartile=0.0663005825517384,
		lower quartile=0.0153468629906844,
		upper whisker=0.142587390653369,
		lower whisker=5.35760026964924e-06,
		draw position=2.3
	},
	] coordinates {};

	\addplot+[
	fill, draw=black,
	boxplot prepared={
		median=0.0136441095873035,
		upper quartile=0.0332003291979788,
		lower quartile=0.00573243411166777,
		upper whisker=0.0743572470432541,
		lower whisker=5.53906518982693e-07,
		draw position=2.7
	},
	] coordinates {};

	\addplot+[
	fill, draw=black,
	boxplot prepared={
		median=0.0156106528398827,
		upper quartile=0.0303397930235736,
		lower quartile=0.00697554100262194,
		upper whisker=0.0653732810401208,
		lower whisker=4.72435078511237e-07,
		draw position=3
	},
	] coordinates {};

	\addplot+[
	fill, draw=black,
	boxplot prepared={
		median=0.0179310823723193,
		upper quartile=0.0343450354699777,
		lower quartile=0.0082245922772126,
		upper whisker=0.0731432995315543,
		lower whisker=3.06074102081461e-06,
		draw position=3.3
	},
	] coordinates {};

	\addplot+[
	fill, draw=black,
	boxplot prepared={
		median=0.00824423041949608,
		upper quartile=0.0214301886080531,
		lower quartile=0.00353805805476005,
		upper whisker=0.0482624271320269,
		lower whisker=1.18492706304886e-06,
		draw position=3.7
	},
	] coordinates {};

	\addplot+[
	fill, draw=black,
	boxplot prepared={
		median=0.00782308111731833,
		upper quartile=0.015144217310251,
		lower quartile=0.00356984614619106,
		upper whisker=0.0324997336055189,
		lower whisker=1.62060317627282e-06,
		draw position=4
	},
	] coordinates {};

	\addplot+[
	fill, draw=black,
	boxplot prepared={
		median=0.00911026710088592,
		upper quartile=0.0176238895739266,
		lower quartile=0.00419236460423904,
		upper whisker=0.0377707995407324,
		lower whisker=1.53910609249033e-06,
		draw position=4.3
	},
	] coordinates {};

	\addplot+[
	fill, draw=black,
	boxplot prepared={
		median=0.00513593170563242,
		upper quartile=0.0153394667136463,
		lower quartile=0.00213494462763337,
		upper whisker=0.0351121238643606,
		lower whisker=5.66020799453576e-07,
		draw position=4.7
	},
	] coordinates {};

	\addplot+[
	fill, draw=black,
	boxplot prepared={
		median=0.00413631131682618,
		upper quartile=0.00819596980337674,
		lower quartile=0.0018960246845309,
		upper whisker=0.0176329804168669,
		lower whisker=1.60746638490201e-07,
		draw position=5
	},
	] coordinates {};

	\addplot+[
	fill, draw=black,
	boxplot prepared={
		median=0.0048696216129997,
		upper quartile=0.00937804875769562,
		lower quartile=0.0022142203859744,
		upper whisker=0.0201192619134435,
		lower whisker=1.33816613822917e-06,
		draw position=5.3
	},
	] coordinates {};

	\addplot+[
	fill, draw=black,
	boxplot prepared={
		median=0.00407951808968225,
		upper quartile=0.0124854012221035,
		lower quartile=0.00158147819241356,
		upper whisker=0.0288197490393539,
		lower whisker=1.64638131557873e-06,
		draw position=5.7
	},
	] coordinates {};

	\addplot+[
	fill, draw=black,
	boxplot prepared={
		median=0.00283216662245692,
		upper quartile=0.00561887446773032,
		lower quartile=0.00129543359513579,
		upper whisker=0.0120631692873489,
		lower whisker=2.54851992328416e-07,
		draw position=6
	},
	] coordinates {};

	\addplot+[
	fill, draw=black,
	boxplot prepared={
		median=0.00343331122672734,
		upper quartile=0.00661674450155594,
		lower quartile=0.00154003299742722,
		upper whisker=0.0142302686654769,
		lower whisker=6.99858913546451e-07,
		draw position=6.3
	},
	] coordinates {};

	\end{axis}
\end{tikzpicture}}\subcaptionbox{$f_{\rm c}=\SI{60}{\giga\hertz}$, $KT=\SI{16}{\milli\second}$, and $S=2$.\protect\label{fig:boxplot_aoa}}[0.33\textwidth]{\begin{tikzpicture}
\definecolor{black28}{RGB}{28,28,28}
\definecolor{brown1814888}{RGB}{181,48,88}
\definecolor{burlywood231175145}{RGB}{231,175,145}
\definecolor{darkgray176}{RGB}{176,176,176}
\definecolor{darkslategray583162}{RGB}{58,31,62}
\definecolor{indianred21811088}{RGB}{218,110,88}
\definecolor{purple1194288}{RGB}{119,42,88}

  \matrix[draw,  inner sep=2pt, fill=white, column sep=0em, ampersand replacement=\&] at (1.95,2.215) {  
        \node [shape=rectangle, draw=black ,fill=purple1194288, label=right:\footnotesize{$\sigma_\alpha=1$°}] {}; \&
        \node [shape=rectangle, draw=black ,fill=brown1814888, label=right:\footnotesize{$\sigma_\alpha=3$°}] {}; \&
        \node [shape=rectangle, draw=black ,fill=indianred21811088, label=right:\footnotesize{$\sigma_\alpha=5$°}] {}; \\
    };
    \tikzset{every mark/.append style={scale=0.9}}
    \matrix[anchor=base, draw,  inner sep=1pt, fill=white] at (3.2,1.4) {  
            \node [shape=diamond, draw=black ,fill=black, scale=0.7, label=right:\tiny{static RX}] {}; \\
    };
  \begin{axis}
    [
    boxplot/draw direction=y,
    ylabel={Error $\varepsilon_{f_{\rm D, t}}$},
    ylabel shift = -5 pt,
    xlabel={SNR [dB]},
    ymin=-0.011, ymax=0.12,
    ytick={0,0.05,0.1,0.15,0.2,0.25,0.3,0.35,0.4,0.45,0.5},
    yticklabels={$0$,$0.05$,$0.1$,,$0.2$,,$0.3$,,$0.4$,,$0.5$},
    yticklabel style = {/pgf/number format/fixed},
    cycle list={{purple1194288},{brown1814888},{indianred21811088}},
    xtick={2,3,4,5},
    ymajorgrids,
    xticklabels={$0$, $5$, $10$, $20$},
    xlabel style={font=\footnotesize}, ylabel style={font=\footnotesize}, ticklabel style={font=\footnotesize},
    /pgfplots/boxplot/box extend=0.25,
    height=3.5cm,
    width=5.5cm,
    ]
	


	\addplot+[
	fill, draw=black,
	boxplot prepared={
		median=0.0154576306992317,
		upper quartile=0.0686669131923394,
		lower quartile=0.00539352872180648,
		upper whisker=0.163500826009092,
		lower whisker=7.0050210773741e-07,
		draw position=1.7
	},
	] coordinates {};
		
	\addplot+[
	fill, draw=black,
	boxplot prepared={
		median=0.0155451234745592,
		upper quartile=0.0682233042649768,
		lower quartile=0.00547051049702431,
		upper whisker=0.162325083363619,
		lower whisker=3.73393479349936e-07,
		draw position=2
	},
	] coordinates {};
		
	\addplot+[
	fill, draw=black,
	boxplot prepared={
		median=0.01672361308297,
		upper quartile=0.0696783429756769,
		lower quartile=0.00586347937895033,
		upper whisker=0.165379007099074,
		lower whisker=5.30904771497057e-06,
		draw position=2.3
	},
	] coordinates {};

	\addplot+[
	fill, draw=black,
	boxplot prepared={
		median=0.00550940749797819,
		upper quartile=0.0150083569393184,
		lower quartile=0.00239475234014292,
		upper whisker=0.0339286925042906,
		lower whisker=9.84102025433151e-07,
		draw position=2.7
	},
	] coordinates {};

	\addplot+[
	fill, draw=black,
	boxplot prepared={
		median=0.00608029616055286,
		upper quartile=0.0169595261585608,
		lower quartile=0.00261099271130958,
		upper whisker=0.0384229413417199,
		lower whisker=6.37243670873998e-07,
		draw position=3
	},
	] coordinates {};

	\addplot+[
	fill, draw=black,
	boxplot prepared={
		median=0.0070367689299547,
		upper quartile=0.0194341803501735,
		lower quartile=0.0030400799993659,
		upper whisker=0.0440209102889859,
		lower whisker=4.76333333926194e-06,
		draw position=3.3
	},
	] coordinates {};

	\addplot+[
	fill, draw=black,
	boxplot prepared={
		median=0.00366952283422067,
		upper quartile=0.00951430308180164,
		lower quartile=0.0015405223683254,
		upper whisker=0.0214593412499629,
		lower whisker=3.82655024734711e-07,
		draw position=3.7
	},
	] coordinates {};

	\addplot+[
	fill, draw=black,
	boxplot prepared={
		median=0.00448092889292549,
		upper quartile=0.0112847390944868,
		lower quartile=0.00190838358082867,
		upper whisker=0.0252834232452499,
		lower whisker=1.02475369679937e-06,
		draw position=4
	},
	] coordinates {};

	\addplot+[
	fill, draw=black,
	boxplot prepared={
		median=0.00550534213180466,
		upper quartile=0.0140270477564603,
		lower quartile=0.00234640682271393,
		upper whisker=0.0315242729144843,
		lower whisker=1.9994498381282e-07,
		draw position=4.3
	},
	] coordinates {};

	\addplot+[
	fill, draw=black,
	boxplot prepared={
		median=0.00234545075188944,
		upper quartile=0.00650597654644193,
		lower quartile=0.000889572422121709,
		upper whisker=0.0149065369894751,
		lower whisker=5.62471484023397e-07,
		draw position=4.7
	},
	] coordinates {};

	\addplot+[
	fill, draw=black,
	boxplot prepared={
		median=0.0034330608628395,
		upper quartile=0.00837384244146687,
		lower quartile=0.00140028815248993,
		upper whisker=0.0188339812680375,
		lower whisker=1.58738690882127e-07,
		draw position=5
	},
	] coordinates {};

	\addplot+[
	fill, draw=black,
	boxplot prepared={
		median=0.00467106542427099,
		upper quartile=0.0110272106448026,
		lower quartile=0.00182340526897922,
		upper whisker=0.0247990181291291,
		lower whisker=1.22537528435736e-07,
		draw position=5.3
	},
	] coordinates {};
 
        \addplot [black, dashed, mark options={solid, fill=black, scale=0.8}, mark=diamond*] table {new_plot/static_rx.dat};
        
 \end{axis}
\end{tikzpicture}}
    \caption{ Target Doppler frequency estimation error varying the number of available static scatterers, the duration of the processing window, the \ac{snr}, and the \acp{aoa} measurements error, with $f_{\rm c}=\SI{60}{\giga\hertz}$, $f_{\rm c}=\SI{28}{\giga\hertz}$, and $f_{\rm c}=\SI{5}{\giga\hertz}$.}
    \vspace{-4mm}
    \label{fig:boxplots}
\end{figure*}

\section{Numerical Results}\label{sec:results}
In this section, we describe our simulations, implemented in Python, and the obtained numerical results.

\vspace{-0.2cm}
\subsection{Simulation setup}\label{sec:sim-setup}

To validate the solution presented in \secref{sec:doppler-est} we perform simulations for $5$~GHz, $28$~GHz, and $60$~GHz carrier frequencies, representing, e.g., IEEE~802.11ax, \ac{fr2} \mbox{5G-NR}, and IEEE~802.11ay systems, respectively. 
\rev{We randomly generate the coordinates of the target and static objects in a $s \times s$ empty mobility area and obtain the real \ac{cir} using \eq{eq:cir-continuous}. Then we simulate the transmission of the waveform used for channel estimation. For $f_c=60$~GHz an IEEE~802.11ay TRN field made of complementary Golay sequences is transmitted, while for $f_c=5$~GHz and $f_c=28$~GHz we use a random sequence of \ac{ofdm} symbols modulated with BPSK. At the receiver, we add \ac{cfo}, \ac{po}, and noise accordingly to the selected \ac{snr} value. Next, the received signal is sampled with period $1/B$ and an estimate of the \ac{cir} is obtained. The phases of the \ac{cir} peaks are given as input to our model.}
All parameters are summarized in \tab{tab:init_param}. \rev{The RX maximum speed, $v_{\rm max}$, differs for each carrier frequency to account for their different typical use case. The maximum phase shift that the Doppler effect can cause is \mbox{$f_{\rm max} = v_{\rm max}/\lambda$}. Finally, we set $T=1/(6f_{\rm max})$ to avoid phase ambiguity as discussed in~\secref{sec:phase-diff}.} \rev{We remark that the selected channel estimation periods are compatible with existing hardware and typical packet transmission times in wireless networks.} We perform $10^4$ simulations for each set of parameters. 
\rev{The residual \ac{cfo} is modeled as $f_{\rm o}(kT) \sim \mathcal{N}(0,\sigma^2_{\rm o})$, where $\sigma^2_{\rm o}$ is chosen such that the frequency shift in $1$~ms is between $\pm 0.1$ parts-per-million~(ppm) of the carrier frequency.}
We model the \ac{po} as $\psi_{\rm o}(kT) \sim \mathcal{N}(0, \sigma^2_{\rm o})$.
Note that, our method cancels out the \ac{cfo} and \ac{po} \textit{exactly}, hence it is not significantly affected by their magnitude.   
The error on the \ac{aoa} is modeled as additive and Gaussian, $\mathcal{N}(0,\sigma_\alpha^2)$.

\begin{table}[t!]
    \centering
    \caption{Simulation parameters used in the numerical results. $\mathcal{U}(a, b)$ denotes the uniform distribution in the interval $[a, b]$. Values in curly brackets refer to the parameters used for $60$, $28$, and $5$~GHz carrier frequencies, respectively.}
    \begin{tabular}{lcc}
       \toprule
       Target Doppler frequency [Hz] & $f_{\rm D, t}$     & $\pm\mathcal{U}(f_{\rm min}, f_{\rm max})$ \\
       Receiver velocity [m/s] & $v^{\rm rx}$       & $\mathcal{U}(v_{\rm min}, v_{\rm max})$ \\
       Min. RX/target Doppler frequency [Hz]& $f_{\rm min}$ & $100$\\
       Min. RX/target velocity [m/s] & $v_{\rm min}$ & $0.5$ \\
       Max. RX/target Doppler frequency [kHz]& $f_{\rm max}$ & \{$1,0.93,0.3$\}\\
       Max. RX/target velocity [m/s] & $v_{\rm max}$ & \{$5,10,20$\} \\ 
       Mobility area side [m] & $s$ & \{$20,50,100$\} \\
       Carrier wavelength [cm] & $\lambda$ & \{$0.5,1.07,6$\}\\
       Channel estimation period [ms] & $T$ & \{$0.166,0.178,0.5$\}\\
       Bandwidth [GHz] & $B$ & \{1.76, 0.4, 0.16\}\\
       Subcarrier spacing [kHz]& $\Delta_f$ & \{$\:\_\:,120,78.125$\} \\  
       \ac{cfo} time difference stand. dev. [MHz]& $\sigma_{\rm o}$ & \{$0.22,0.12,0.02$\}\\
       \bottomrule
    \end{tabular}
    \label{tab:init_param}
\end{table}

\vspace{-0.2cm}
\subsection{Doppler frequency estimation performance}
We evaluate the performance of our Doppler frequency estimation algorithm in terms of normalized absolute estimation error, defined as $\varepsilon_{f_{\rm D, t}} = |f_{\rm D, t} - \hat{f}_{\rm D, t}| / |f_{\rm D, t}|$.

\subsubsection{Number of static scatterers}
\label{sec:npath}
\rev{The number of available static scatterers, $S$, is key in determining the algorithm's performance. As detailed in \secref{sec:nls-sol}, the target Doppler frequency can be estimated as long as at least $2$ static scatterers are resolved at the RX. In \fig{fig:boxplot_path}, we show that the median estimation error and its spread are reduced if more static scatterers are available, however beyond $S=4$ the performance improvement is limited. The median error lies below $2$\% of the actual Doppler frequency even with $S=2$ and for every considered carrier frequency. Each scatterer adds one equation to the \ac{nls} problem in \secref{sec:nls-sol} without increasing the number of unknowns, leading to a more robust and, on the other hand, more complex to compute solution. In \fig{fig:boxplot_path}, black diamonds represent the average time (across all $f_{\rm c}$) needed to solve the \ac{nls} problem on an Intel Core i7-9700 CPU, by varying $S$.}

\subsubsection{Processing window duration}
\label{sec:res_interval}
\rev{In \fig{fig:boxplot_K}, we evaluate the impact of averaging phase measurements over a longer processing window. Increasing the number of considered frames $K$, while keeping $T$ constant, clearly improves robustness to noise. In addition, we highlight that since $T$ differs for each carrier frequency, for a fixed time interval $KT$ every considered $f_{\rm c}$ entails a different $K$.}
However, the window duration can not be increased arbitrarily but, has to be tuned depending on the dynamicity of the multipath environment to ensure that the parameters remain constant throughout the processing window. 
\rev{As an example, for indoor human sensing applications, the movement velocities involved could be considered constant up to a few tens of milliseconds. With $KT=\SI{32}{\milli\second}$, our method provides a median error below $1$\% of the true Doppler frequency for every considered $f_{\rm c}$.}

\subsubsection{Measurements error}
\label{sec:res_snr}
In \fig{fig:boxplot_aoa}, we show the Doppler frequency estimation error depending on the \ac{snr}, and the noise affecting \ac{aoa} measurements ($\sigma_{\alpha}$). 
For lower \ac{snr} values, the estimation becomes noisy, despite the median normalized error below $0.02$. This is due to the required trade-off in the choice of the \ac{cir} measurement interval $T$, which is discussed in the next section. \fig{fig:boxplot_aoa} also shows that the Doppler frequency estimate is slightly affected by the \ac{aoa} error, which means the \ac{snr} level has a stronger impact. 
\rev{As a reference for comparison, we report the median estimation error on $f_{\rm D,t}$ with a \textit{static} RX. In this case, $f_{\rm D,t}$ is estimated directly after \eq{eq:phi-trg}, hence it is not affected by the \ac{aoa} estimation errors, but it is still impacted by noise.} 
\rev{Although our work focuses on Doppler frequency estimation, the proposed solution also provides an estimate of $\eta$ and $v^{\rm rx}$. In \fig{fig:speed_eta} we provide the normalized estimation errors for $\eta$ and $v^{\rm rx}$, respectively \mbox{$\varepsilon_\eta = |\eta - \hat{\eta}| / |\eta|$} and \mbox{$\varepsilon_{v^{\rm rx}}=|v^{\rm rx} - \hat{v}^{\rm rx}| / |v^{\rm rx}|$}, where $v^{\rm rx}$ is the estimated RX speed. The estimates of $\eta$, and especially of $v^{\rm rx}$, are less accurate than that of the Doppler frequency and would only be useful for very high \acs{snr}. These results could be improved by using additional information from an onboard motion sensor at the RX.}

\subsubsection{CIR measurement interval analysis} 
\label{sec:res_T}
\rev{As discussed in \secref{sec:phase-diff}, $T$ has to be sufficiently small to avoid ambiguity in the phase measurements. However, small values of $T$ make the estimation of the Doppler frequency more sensitive to noise due to the structure of \eq{eq:fD}. 
For each $f_{\rm c}$, we numerically evaluate the impact of increasing $T$, starting from the maximum value for which ambiguities are avoided. We observe that increasing $T$ causes a linear growth in $\varepsilon_{f_{\rm D, t}}$ due to phase ambiguities. Taking as an example $f_{\rm c}=\SI{28}{\giga\hertz}$, the maximum channel estimation period used in the simulations is $T=\SI{0.178}{\milli\second}$ and increasing it to $T=\SI{0.28}{\milli\second}$ leads to a normalized error on the Doppler estimate of $\varepsilon_{f_{\rm D, t}}>0.5$.} 


\begin{figure}
    \centering
    \begin{subfigure}{\columnwidth}
        \centering
        \hspace{1cm}
        \begin{tikzpicture}
\definecolor{black28}{RGB}{28,28,28}
\definecolor{brown1814888}{RGB}{181,48,88}
\definecolor{burlywood231175145}{RGB}{231,175,145}
\definecolor{darkgray176}{RGB}{176,176,176}
\definecolor{darkslategray583162}{RGB}{58,31,62}
\definecolor{indianred21811088}{RGB}{218,110,88}
\definecolor{purple1194288}{RGB}{119,42,88}
\matrix[draw, fill=white, ampersand replacement=\&, 
inner sep=2pt, 
    ] at (1.6,2.8) {  
        \node [shape=rectangle, draw=black ,fill=purple1194288, label=right:\scriptsize{$f_{\rm c}=\SI{60}{\giga\hertz}$}] {};\&
        \node [shape=rectangle, draw=black ,fill=brown1814888, label=right:\scriptsize{$f_{\rm c}=\SI{28}{\giga\hertz}$}] {}; \&
        \node [shape=rectangle, draw=black ,fill=indianred21811088, label=right:\scriptsize{$f_{\rm c}=\SI{5}{\giga\hertz}$}] {};\\
    };
\end{tikzpicture}
        \vspace{0.05cm}
    \end{subfigure}
    \begin{subfigure}{0.49\columnwidth}
        \centering
        \begin{tikzpicture}
\definecolor{black28}{RGB}{28,28,28}
\definecolor{brown1814888}{RGB}{181,48,88}
\definecolor{burlywood231175145}{RGB}{231,175,145}
\definecolor{darkgray176}{RGB}{176,176,176}
\definecolor{darkslategray583162}{RGB}{58,31,62}
\definecolor{indianred21811088}{RGB}{218,110,88}
\definecolor{purple1194288}{RGB}{119,42,88}
  \begin{axis}
    [
    boxplot/draw direction=y,
    ylabel={Error $\varepsilon_\eta$},
    xlabel shift = -1 pt,
    ylabel shift = -5 pt,
    xlabel={SNR [dB]},
    ymin=-0.08, ymax=0.85,
    ytick={0,0.25,0.5,0.75,1,1.25,1.5,1.75,2},
    yticklabels={$0$,$0.25$,$0.5$,$0.75$,$1$,,$1.5$,,$2$},
    yticklabel style = {/pgf/number format/fixed},
    cycle list={{purple1194288},{brown1814888},{indianred21811088}},
    xtick={1,2,3,4},
    ymajorgrids,
    xticklabels={ $0$, $5$, $10$, $20$},
    xlabel style={font=\footnotesize}, ylabel style={font=\footnotesize}, ticklabel style={font=\footnotesize},
    /pgfplots/boxplot/box extend=0.25,
    height=3.5cm,
    width=4.8cm,
    ]

    \addplot+[
	fill, draw=black,
	boxplot prepared={
		median=0.0743237695953792,
		upper quartile=0.206135291606201,
		lower quartile=0.0250469418675565,
		upper whisker=0.476396726237431,
		lower whisker=1.79397713176381e-05,
		draw position=0.7
	},
	] coordinates {};

	\addplot+[
	fill, draw=black,
	boxplot prepared={
		median=0.104718867510952,
		upper quartile=0.259587035126255,
		lower quartile=0.0360023918413062,
		upper whisker=0.594728137478109,
		lower whisker=1.96737785486505e-05,
		draw position=1
	},
	] coordinates {};

	\addplot+[
	fill, draw=black,
	boxplot prepared={
		median=0.0942580057880106,
		upper quartile=0.221637987450462,
		lower quartile=0.038496562844569,
		upper whisker=0.495060916712417,
		lower whisker=3.04886629104449e-05,
		draw position=1.3
	},
	] coordinates {};

	\addplot+[
	fill, draw=black,
	boxplot prepared={
		median=0.0450211177596119,
		upper quartile=0.125664403392813,
		lower quartile=0.0161710811733704,
		upper whisker=0.289850707488633,
		lower whisker=9.76869410612856e-06,
		draw position=1.7
	},
	] coordinates {};

	\addplot+[
	fill, draw=black,
	boxplot prepared={
		median=0.0371555785719232,
		upper quartile=0.0867804070636233,
		lower quartile=0.0148192363785734,
		upper whisker=0.19449449517215,
		lower whisker=1.07370627173695e-06,
		draw position=2
	},
	] coordinates {};

	\addplot+[
	fill, draw=black,
	boxplot prepared={
		median=0.0648416570197363,
		upper quartile=0.149927138937161,
		lower quartile=0.0267686447896065,
		upper whisker=0.334344758711311,
		lower whisker=6.92600327702014e-06,
		draw position=2.3
	},
	] coordinates {};

	\addplot+[
	fill, draw=black,
	boxplot prepared={
		median=0.0374447116175446,
		upper quartile=0.098088254183698,
		lower quartile=0.013801602610322,
		upper whisker=0.224253179369207,
		lower whisker=1.20176316812749e-06,
		draw position=2.7
	},
	] coordinates {};

	\addplot+[
	fill, draw=black,
	boxplot prepared={
		median=0.030360415596439,
		upper quartile=0.0719903182875164,
		lower quartile=0.0117818726118621,
		upper whisker=0.162137958831305,
		lower whisker=9.44550295797736e-06,
		draw position=3
	},
	] coordinates {};

	\addplot+[
	fill, draw=black,
	boxplot prepared={
		median=0.0505060459984735,
		upper quartile=0.116677070241504,
		lower quartile=0.0201921765211281,
		upper whisker=0.261360984284302,
		lower whisker=3.55916375963035e-07,
		draw position=3.3
	},
	] coordinates {};

	\addplot+[
	fill, draw=black,
	boxplot prepared={
		median=0.0322717892985155,
		upper quartile=0.0825059678033986,
		lower quartile=0.0119252386855119,
		upper whisker=0.188339290244024,
		lower whisker=3.12479776449224e-06,
		draw position=3.7
	},
	] coordinates {};

	\addplot+[
	fill, draw=black,
	boxplot prepared={
		median=0.0248033436345417,
		upper quartile=0.0607037407092333,
		lower quartile=0.00971105728880997,
		upper whisker=0.13704325543688,
		lower whisker=3.24329299468213e-06,
		draw position=4
	},
	] coordinates {};

	\addplot+[
	fill, draw=black,
	boxplot prepared={
		median=0.0414343347344155,
		upper quartile=0.0975664981202827,
		lower quartile=0.016023747407107,
		upper whisker=0.219831707819459,
		lower whisker=7.01273735497963e-06,
		draw position=4.3
	},
	] coordinates {};

    \end{axis}
\end{tikzpicture}
        \vspace{-.2cm}
        \caption{Normalized $\eta$ estimation error.}
        \label{fig:boxplot_eta}
    \end{subfigure}
    \begin{subfigure}{0.49\columnwidth}
        \centering
        \begin{tikzpicture}
\definecolor{black28}{RGB}{28,28,28}
\definecolor{brown1814888}{RGB}{181,48,88}
\definecolor{burlywood231175145}{RGB}{231,175,145}
\definecolor{darkgray176}{RGB}{176,176,176}
\definecolor{darkslategray583162}{RGB}{58,31,62}
\definecolor{indianred21811088}{RGB}{218,110,88}
\definecolor{purple1194288}{RGB}{119,42,88}
  \begin{axis}
    [
    boxplot/draw direction=y,
    ylabel={Error $\varepsilon_{v^{\rm rx}}$},
    xlabel shift = -1 pt,
    ylabel shift = -5 pt,
    xlabel={SNR [dB]},
    ymin=-0.08, ymax=0.85,
    ytick={0,0.25,0.5,0.75,1,1.25,1.5,1.75,2},
    yticklabels={$0$,$0.25$,$0.5$,$0.75$,$1$,,$1.5$,,$2$},
    yticklabel style = {/pgf/number format/fixed},
    cycle list={{purple1194288},{brown1814888},{indianred21811088}},
    xtick={1,2,3,4},
    ymajorgrids,
    xticklabels={ $0$, $5$, $10$, $20$},
    xlabel style={font=\footnotesize}, ylabel style={font=\footnotesize}, ticklabel style={font=\footnotesize},
    /pgfplots/boxplot/box extend=0.25,
    height=3.5cm,
    width=4.8cm,
    ]

    \addplot+[
	fill, draw=black,
	boxplot prepared={
		median=0.188691693833259,
		upper quartile=0.520221750434927,
		lower quartile=0.0656593819994641,
		upper whisker=1.20158151962962,
		lower whisker=7.28762175074445e-06,
		draw position=0.7
	},
	] coordinates {};

	\addplot+[
	fill, draw=black,
	boxplot prepared={
		median=0.25891764859988,
		upper quartile=0.697278552993303,
		lower quartile=0.090058520010702,
		upper whisker=1.60568526689773,
		lower whisker=1.58326982264409e-05,
		draw position=1
	},
	] coordinates {};

	\addplot+[
	fill, draw=black,
	boxplot prepared={
		median=0.261218211827209,
		upper quartile=0.581835341270658,
		lower quartile=0.0981970376574996,
		upper whisker=1.30670208467602,
		lower whisker=3.23045143325525e-05,
		draw position=1.3
	},
	] coordinates {};

	\addplot+[
	fill, draw=black,
	boxplot prepared={
		median=0.116652361249008,
		upper quartile=0.292221594396714,
		lower quartile=0.0439963542902754,
		upper whisker=0.664555976993339,
		lower whisker=2.19136218210972e-05,
		draw position=1.7
	},
	] coordinates {};

	\addplot+[
	fill, draw=black,
	boxplot prepared={
		median=0.100717020816473,
		upper quartile=0.22236355985569,
		lower quartile=0.0390286392914756,
		upper whisker=0.497118654205084,
		lower whisker=5.54532322171233e-05,
		draw position=2
	},
	] coordinates {};

	\addplot+[
	fill, draw=black,
	boxplot prepared={
		median=0.179169489485891,
		upper quartile=0.385856003745102,
		lower quartile=0.0694741839614141,
		upper whisker=0.860030243207658,
		lower whisker=6.23301227537737e-06,
		draw position=2.3
	},
	] coordinates {};

	\addplot+[
	fill, draw=black,
	boxplot prepared={
		median=0.0982604703720517,
		upper quartile=0.239153555825274,
		lower quartile=0.0369097145560284,
		upper whisker=0.541272642526236,
		lower whisker=2.05017053564481e-05,
		draw position=2.7
	},
	] coordinates {};

	\addplot+[
	fill, draw=black,
	boxplot prepared={
		median=0.0802448453052423,
		upper quartile=0.1754563387893,
		lower quartile=0.0318808510116896,
		upper whisker=0.390177178492074,
		lower whisker=2.86110869132814e-06,
		draw position=3
	},
	] coordinates {};

	\addplot+[
	fill, draw=black,
	boxplot prepared={
		median=0.135548558145441,
		upper quartile=0.292951385964636,
		lower quartile=0.053401015624624,
		upper whisker=0.651956278718631,
		lower whisker=3.63832352054051e-05,
		draw position=3.3
	},
	] coordinates {};

	\addplot+[
	fill, draw=black,
	boxplot prepared={
		median=0.0836460210135478,
		upper quartile=0.202469361502368,
		lower quartile=0.0321212294719916,
		upper whisker=0.456956976195309,
		lower whisker=1.85228097169358e-05,
		draw position=3.7
	},
	] coordinates {};

	\addplot+[
	fill, draw=black,
	boxplot prepared={
		median=0.0658882441645434,
		upper quartile=0.147190639209893,
		lower quartile=0.0262510125694735,
		upper whisker=0.327799029499976,
		lower whisker=6.27494561863599e-06,
		draw position=4
	},
	] coordinates {};

	\addplot+[
	fill, draw=black,
	boxplot prepared={
		median=0.106344005316074,
		upper quartile=0.240011265894288,
		lower quartile=0.0435928831430895,
		upper whisker=0.534398187438831,
		lower whisker=8.40985740744362e-06,
		draw position=4.3
	},
	] coordinates {};
 	\end{axis}
\end{tikzpicture}
        \vspace{-.61cm}
        \caption{Normalized $v^{\rm rx}$ estimation error.}
        \label{fig:boxplot_speed}
    \end{subfigure}
    \caption{Varying \ac{snr} values, with $\sigma_\alpha=5$°, $KT=\SI{16}{\milli\second}$, and $S=2$.}
    \label{fig:speed_eta}
\end{figure}

\section{Conclusion}\label{sec:conclusion}

In this letter, we proposed the first method to estimate the Doppler frequency of a target in an \textit{asynchronous} \ac{isac} system with \textit{mobile} devices. Our approach can effectively disentangle the target's Doppler frequency from the \ac{cfo} and the Doppler caused by the device movement. It does so by leveraging (i)~phase differences across multipath components and time, and (ii)~the multipath geometry. The Doppler frequency is estimated by solving an \ac{nls} problem that requires the \ac{los} and at least $2$ reflections from static scatterers to be available. 

Our simulation results show that the proposed method achieves accurate Doppler frequency estimation at medium and high \ac{snr}, and it is robust to \ac{aoa} estimation errors. 

\bibliography{references.bib}
\bibliographystyle{IEEEtran}
\end{document}